\documentclass{webofc}

\usepackage[varg]{txfonts}   
\usepackage{hyperref}
\usepackage{url}
\usepackage{subfigure}
\usepackage{lineno}
\hypersetup{colorlinks=true,citecolor=blue,urlcolor=blue,linkcolor=blue}
\begin{document}
\title{Exploring Collective Effects in Small Collision Systems via Light-Flavour Particle Production in High-Multiplicity pp Collisions at $\sqrt{s}=$ 13 TeV in ALICE}
%
%
\author{\firstname{Rajendra Nath} \lastname{Patra}\inst{1}\fnsep\thanks{\email{rajendra.nath.patra@cern.ch, rajendrapatra07@gmail.com}}
	\and
	\firstname{} \lastname{for the ALICE Collaboration}
}

\institute{
           University of Jammu, Jammu, India 
          }

\abstract{High-multiplicity pp and p--Pb collisions show signatures of radial and anisotropic flow, as well as the enhanced production of strange hadrons. These observations resemble phenomena seen in heavy-ion collisions, where a quark–gluon plasma (QGP) is formed, although their origin in small systems remains under discussion. This contribution presents ALICE Run 2 results on light-flavour hadron ($\pi$, K, p) production in high-multiplicity (HM) pp collisions at $\sqrt{s}=$ 13 TeV, where the charged-particle multiplicity reaches values comparable to peripheral or semi-peripheral Pb–Pb collisions.
The multiplicity dependence of $\pi$, K, and p transverse momentum spectra, $p_{\rm T}$-differential and $p_{\rm T}$-integrated particle ratios, and mean transverse momenta ($\langle p_{\rm T} \rangle$) is presented and compared with published results from other collision systems and energies, as well as with PYTHIA 8 and EPOS4 model calculations.
}

\maketitle
\section{Introduction}
\label{intro}
Ultra-relativistic heavy-ion collisions create a hot and dense medium of deconfined quarks and gluons, the quark–gluon plasma (QGP)~\cite{Burza_2018}, which has been reported by the Relativistic Heavy Ion Collider (RHIC) and the Large Hadron Collider (LHC) through different types of studies in light flavour sectors, like double-ridge structure in particle correlation studies, anisotropic flow $v_2$ coefficient, characteristic increment of baryon-to-meson ratio, and enhancement of strangeness production, which explains the collective nature of the QGP medium~\cite{ALICE_QCD_2024,STAR_QGP_2005}. The measurements in pp and p--Pb collision systems have also reported multiplicity dependent double-ridge structure in long-range correlation and non-zero  $v_2$ coefficient~\cite{CMS_pp_13TeV_2017_v2, CMS_longrange_2010, ATLAS_pp_2016, ALICE_flow_pp_2026}. The strangeness enhancement is one of the earliest signatures of quark–gluon plasma (QGP) formation in relativistic heavy-ion collisions. The strangeness production study among different collision systems and collision energies presents a smooth increment with final-state charged-particle multiplicity density~\cite{ALICE_strange_Nature_2017, ALICE:2019avo, ALICE_QCD_2024}.

The above-mentioned observations in small systems (pp, p--Pb) require a comprehensive explanation of the mechanisms behind these phenomena. This work presents the latest measurements from LHC Run 2 data on light-flavour production in high-multiplicity pp collisions at $\sqrt{s}$ = 13 TeV~\cite{ALICE_pp_13TeV_HM_2026} in the rapidity region $|y|<0.5$. The high-multiplicity triggered pp collisions can play a significant role, as the charged-particle density of such events is comparable to the peripheral Pb--Pb collisions. We will present the transverse-momentum spectra of $\pi$, K, and p with different multiplicity classes, mean transverse momenta $\langle p_{\rm T} \rangle$, particle yield ratio, and the ratio of $p_{\rm T}$-spectra of kaon-to-pion and proton-to-pion. The particle ratios are compared with the published V0M I and minimum-bias (MB) multiplicity classes in pp collisions at $\sqrt{s}$ = 13 TeV \cite{pp_13TeV_2020, pp_7_13_TeV_MB_2021}.
The new measurements are compared with results from p--Pb and Pb--Pb collision systems at different collision energies available at the LHC. Such comparisons provide an opportunity to investigate collective-like effects in the pp system and to assess the sensitivity of the observables to different theoretical model descriptions. In particular, we consider PYTHIA 8~\cite{Pythia8_2008}, a Monte Carlo event generator based on the Lund string fragmentation model, which can describe high-multiplicity pp events through color-rope fragmentation mechanisms, enhanced strangeness production, and color reconnection. We also consider the EPOS4 model~\cite{EPOS4_2023, EPOS4_2024}, which employs a core--corona approach. In this framework, high-density string segments form the core, which undergoes collective expansion, while low-density string segments constitute the corona and hadronize through string fragmentation. The core provides the initial conditions for the subsequent hydrodynamic evolution, which is applicable to collision systems ranging from pp to heavy-ion (AA) collisions. In contrast, the corona component contributes predominantly through string fragmentation and becomes increasingly relevant toward lower multiplicities in pp collisions. A detailed comparison of the measurements with the model predictions is presented and discussed throughout the paper.

\section{Results}
A detailed description of the ALICE detector can be found in Ref.~\cite{ALICE_perf_2008}. The sub-detectors used in this analysis are the V0M (combined of V0A and V0C detectors), the Inner Tracking System (ITS), the Time Projection Chamber (TPC), and the Time-Of-Flight (TOF) [47]. The analyzed data correspond to pp collisions at $\sqrt{s}=$ 13 TeV collected by the ALICE experiment during the LHC Run 2 in 2016, with an integrated luminosity of 13~pb$^{-1}$ for the high-multiplicity (HM) triggered sample. The HM-triggered events are selected using a threshold on the V0M amplitude corresponding to a charged-particle density of ${\rm d}N_{\rm ch}/{\rm d}\eta > 30.1$, defining the top 0.1\% of the MB event sample, for which the equivalent ${\rm d}N_{\rm ch}/{\rm d}\eta$ is 6.93 \cite{pp_13TeV_multiplicity_2021}.

\begin{figure}[h]
	\centering
		\includegraphics[width=1.0\textwidth]{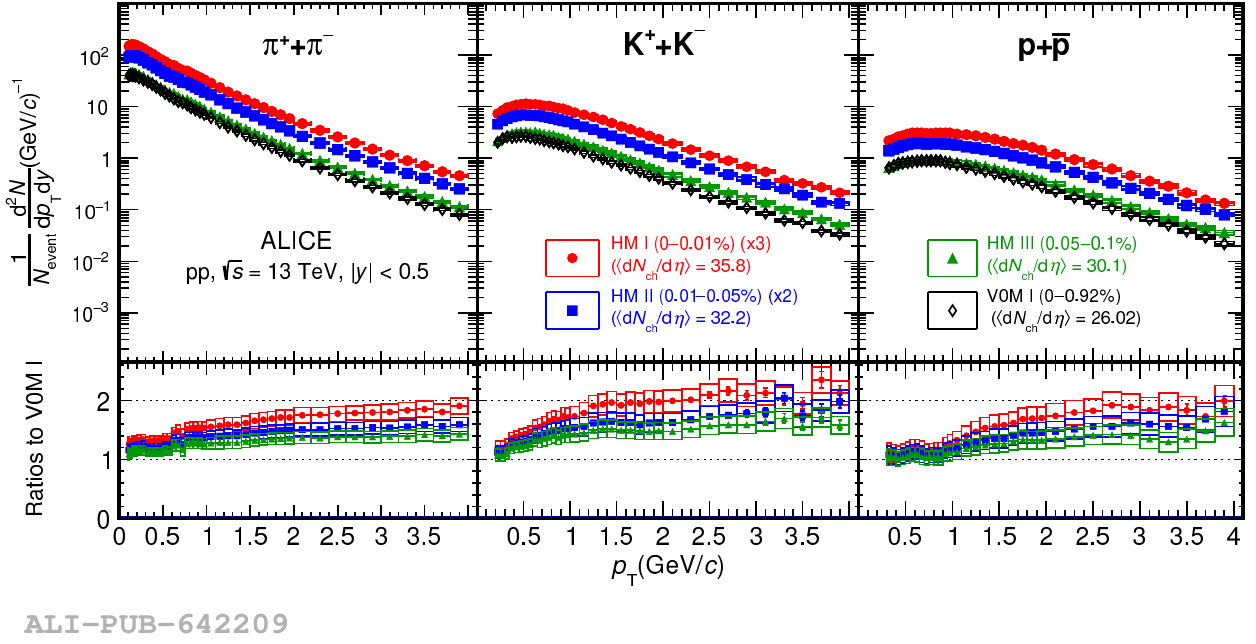}
	\caption{$p_{\rm T}$-differential spectra of $\pi$, K and p for different high-multiplicity classes. The spectra are scaled to improve visibility. The ratios to V0M I spectra~\cite{pp_13TeV_2020} are shown in the bottom panels. Vertical bars and open boxes represent the statistical and systematic uncertainties, respectively.}
	\label{fig:pi_spectra}
\end{figure}

The $p_{\rm T}$-differential spectra of the $\pi$, K and p are measured for three high-multiplicity classes defined as, HM I, HM II and HM III are corresponding to the 0-0.01\%, 0.01-0.05\% and 0.05-0.1\%, respectively, of pp collisions at $\sqrt{s}$ = 13~TeV are shown in Fig.~\ref{fig:pi_spectra}. The results are compared with the MB events~\cite{ALICE_pp_13TeV_HM_2026} along with the reference V0M I ($0-0.92\%$) multiplicity class measured by ALICE~\cite{pp_13TeV_2020} at the same collision system and energy. The average charged-particle multiplicity density $\langle {\rm d} N_{\rm {ch}}/\rm{d}\eta\rangle$, related to each HM class, is also displayed in the figure. The bottom panels of the figure show the ratios of the HM spectra to the V0M I class \cite{pp_13TeV_2020}. A multiplicity-dependent hardening of the spectra is observed, with a stronger effect for heavier particles. Similar patterns have been observed in other collision systems and are consistent with radial-flow effects, although alternative mechanisms can also contribute in small systems\cite{pp_13TeV_2020, pPb_5020GeV_2014, PbPb_5020GeV_2020}. 

\begin{figure}[h]
	\centering
	\includegraphics[width=1.0\textwidth]{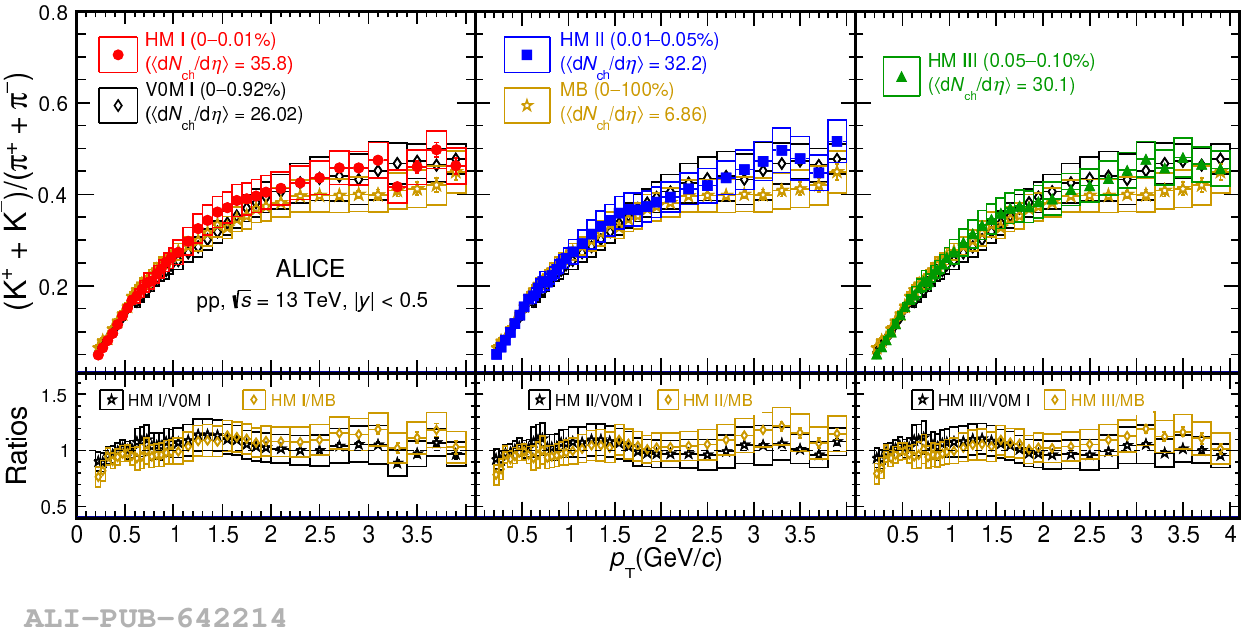}
	
	\caption{(Top panels) Ratios ${\rm K}/{\pi}$ of the $p_{\rm T}$-spectra for all high-multiplicity classes, V0M I and minimum-bias results \cite{pp_13TeV_2020, pp_7_13_TeV_MB_2021}. (Bottom panels) Ratios to V0M I and minimum-bias classes. The boxes represent the systematic uncertainties, while the bars represent the statistical ones.}
	\label{fig:ratio_ktopi_pT}
\end{figure}

Figure~\ref{fig:ratio_ktopi_pT} presents the $p_{\rm T}$-differential ${\rm K}/{\pi}$ ratios for the different multiplicity classes. The ${\rm K}/{\pi}$ ratio exhibits an increasing trend at low $p_{\rm T}$ and saturates at higher $p_{\rm T}$.  The bottom panel shows the ratios of the HM results with the V0M I and MB classes. The HM results are consistent with those of V0M I and MB classes and show no dependence on multiplicity.

\begin{figure}[h]
	\centering
	\includegraphics[width=1.0\textwidth]{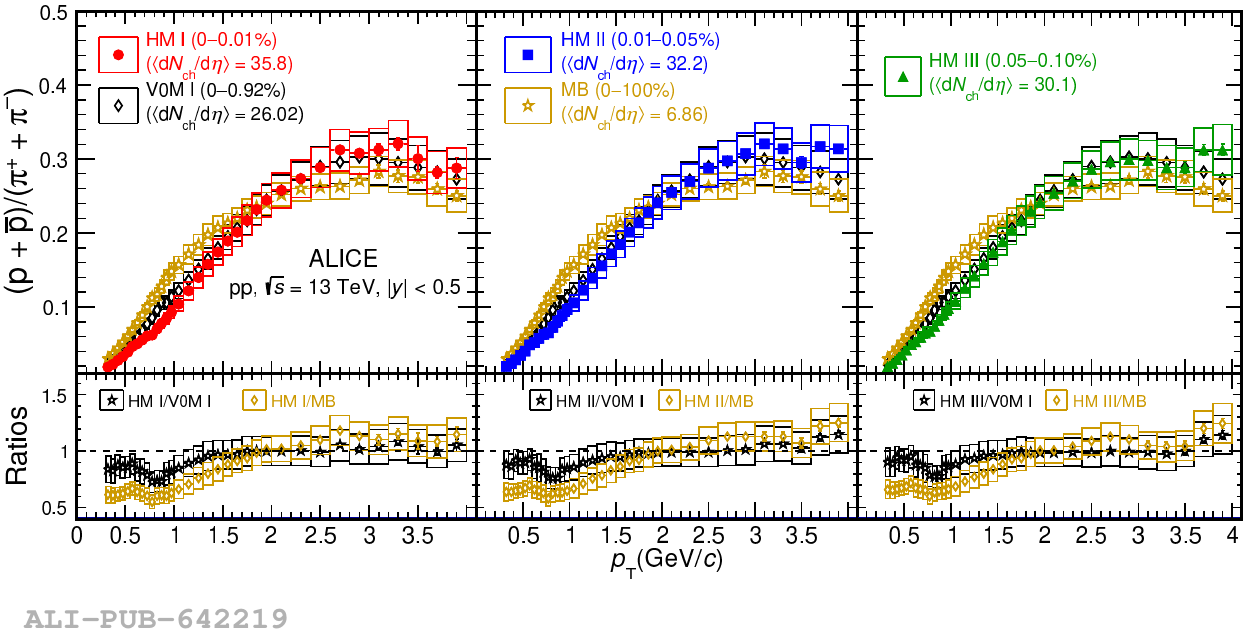}
	
	\caption{(Top panels) Ratios ${\rm p}/{\pi}$ of the $p_{\rm T}$-spectra for all high-multiplicity classes, V0M I and minimum-bias results \cite{pp_13TeV_2020, pp_7_13_TeV_MB_2021}. (Bottom panels) Ratios to V0M I and minimum-bias classes. The boxes represent the systematic uncertainties, while the bars represent the statistical ones.}
	\label{fig:ratio_ptopi_pT}
\end{figure} 

Figure~\ref{fig:ratio_ptopi_pT} presents the $p_{\rm T}$-differential ${\rm p}/{\pi}$ ratios for the different multiplicity classes. 
However, the ${\rm p}/{\pi}$ ratio exhibits a suppression at low $p_{\rm T}$ ($<1~{\rm GeV}/c$) and an enhancement at intermediate $p_{\rm T}$ ($\sim3~{\rm GeV}/c$). The bottom panels of the figure represent the ratio of the HM results with respect to the V0M I and MB multiplicity classes. It shows a systematic increasing trend  from HM I to HM III; however, the large uncertainties limit the significance of this trend. A similar enhancement of ${\rm p}/{\pi}$ ratio has been reported at intermediate $p_{\rm T}$ in collision systems p--Pb and Pb--Pb, where it is attributed to radial flow of quarks in the medium that boosts heavier hadrons at intermediate $p_{\rm T}$~\cite{pPb_5020GeV_2014, PbPb_2760GeV_2013, PbPb_5020GeV_2020}.

\begin{figure}[h]
	\centering
	\vspace{-3 mm}
	\includegraphics[width=1\textwidth]{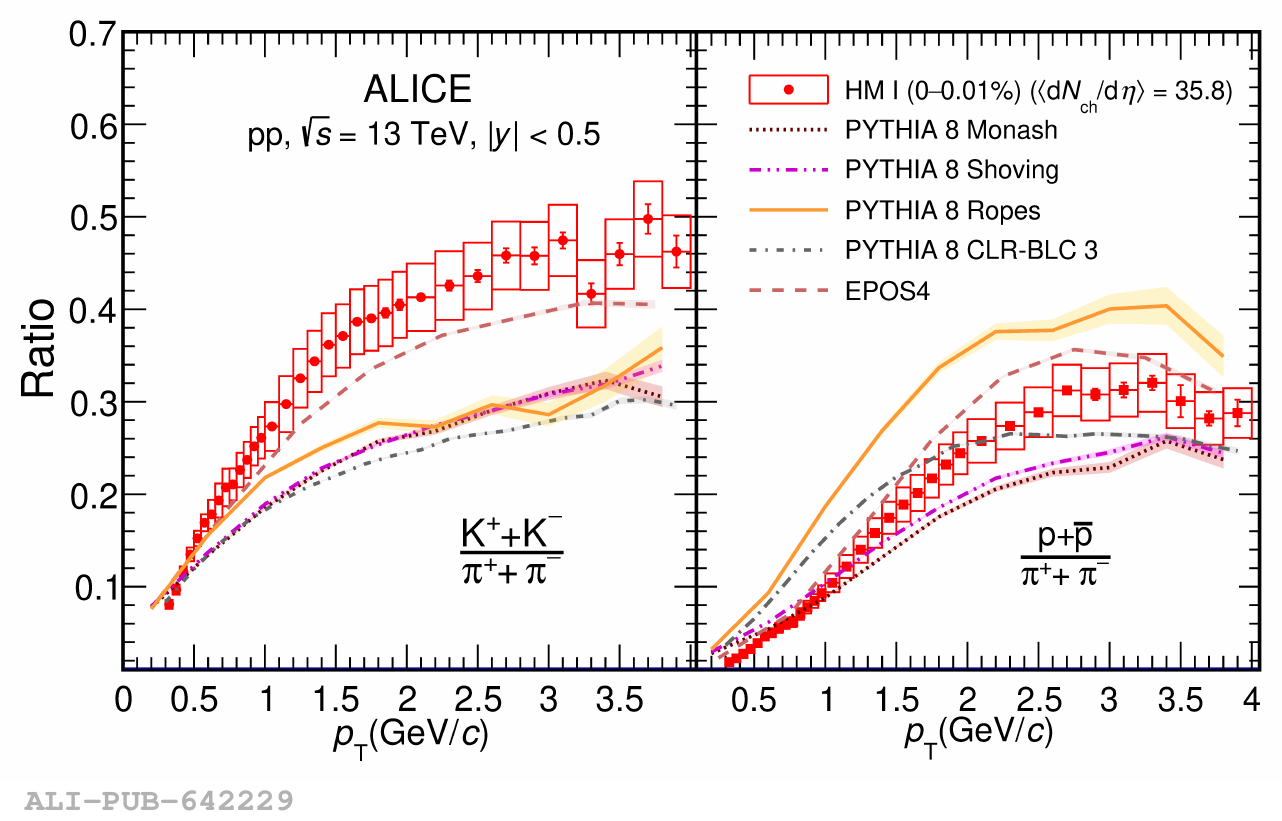}
	
	\caption{Ratios ${\rm K}/{\pi}$ (left) and ${\rm p}/{\pi}$ (right) of the $p_{\rm T}$-spectra compared with PYTHIA 8: Monash, Shoving, Ropes, CLR-BLC 3 tunes, and EPOS4  MC event generator for the HM I multiplicity class. The empty boxes are the systematic uncertainties of reported spectra ratios, and the bands represent the statistical uncertainties in the models.}
	\label{fig:ratio_ktopi_ptopi_pT_model}
\end{figure}

The measured results are compared with predictions from the PYTHIA 8 and EPOS4 Monte Carlo event generators. The PYTHIA 8 Monash 2013 tune includes multiparton interactions (MPI) and color-reconnection mechanisms. In the PYTHIA 8 Rope model, overlapping strings can lead to coherent interactions among the color charges at string endpoints and at gluon kinks, resulting in the formation of a color rope that is hadronized with an effectively larger string tension. The PYTHIA 8 string-shoving mechanism allows strings to interact through a repulsive force before hadronization, generating an effective transverse microscopic pressure that can contribute to long-range correlations. The PYTHIA 8 CLR-BLC model is designed to describe enhanced baryon production during string fragmentation. This model includes different modes; for the present analysis, mode 3 is used for the simulations.
The EPOS4 Monte Carlo model provides a framework for describing high-energy proton--proton and nucleus--nucleus collisions, incorporating multiple scattering and saturation effects. The initial conditions in EPOS4 are described in terms of flux tubes within the Gribov--Regge multiple-scattering framework, followed by hydrodynamic evolution of the core component in the core--corona approach. The measured ${\rm K}/{\pi}$ and ${\rm p}/{\pi}$ ratios are compared with predictions from different PYTHIA 8 tunes and EPOS4, as shown in Fig.~\ref{fig:ratio_ktopi_ptopi_pT_model}. The PYTHIA 8 predictions generally show deviations from the data for both ratios, with the exception of the PYTHIA 8 CLR-BLC mode 3, which provides an improved qualitative description of the ${\rm p}/{\pi}$ ratio. The EPOS4 predictions provide a comparatively better description of the data for both the ${\rm K}/{\pi}$ and ${\rm p}/{\pi}$ ratios than the considered PYTHIA 8 tunes.

Integrated results of $p_{\rm T}$ spectra are used to calculate $p_{\rm T}$-integrated particle yield $\rm{d}N/\rm{d}y$ and mean transverse momentum $\langle p_{\rm T} \rangle$ for all three HM-classes. A Lévy-Tsallis function fit is used to extrapolate the contribution from outside of the $p_{\rm T}$ range of this analysis. 
A number of fit functions, namely Bose-Einstein, Fermi-Dirac, $p_{\rm T}$ exponential and Blast-Wave are used to estimate the systematic uncertainty in the $p_{\rm T}$-integrated measurements.

\begin{figure}[h]
	\centering
	\includegraphics[width=0.7\textwidth]{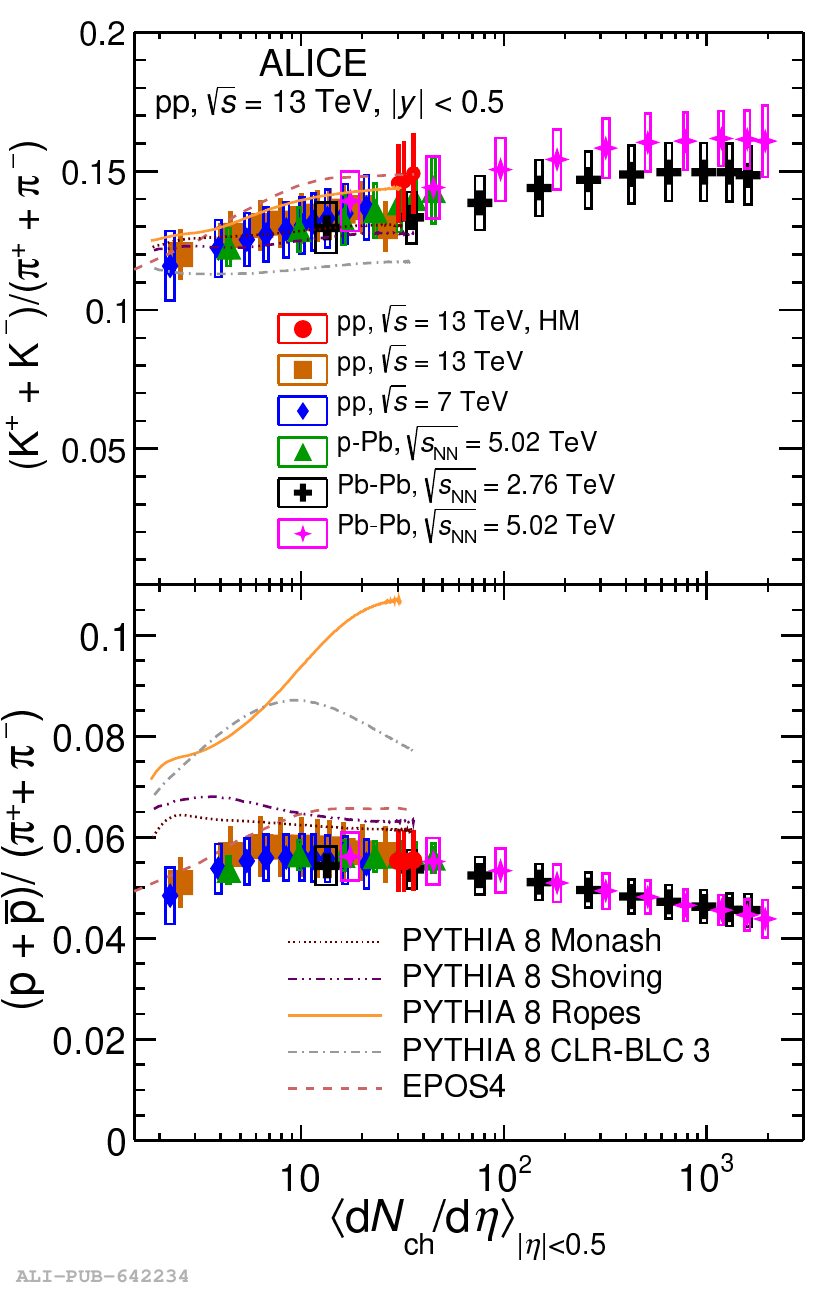}	
	\caption{${\rm K}/{\pi}$ (top) and ${\rm p}/{\pi}$ (bottom) $p_{\rm T}$-integrated yield ratios as a function of charged-particle multiplicity density ($\langle {\rm d} N_{\rm ch}/{\rm d} \eta \rangle$) for high-multiplicity pp collisions at $\sqrt{s}$ = 13 TeV. The results are compared with pp, p--Pb and Pb--Pb collisions at different collision energies \cite{pp_13TeV_2020, pp_7_TeV_2019, pPb_5020GeV_2014, PbPb_2760GeV_2013,  PbPb_5020GeV_2020}. Empty boxes represent the total systematic uncertainties and bars indicate the statistical ones. Colored lines represent the predictions from different tunes of PYTHIA 8 and EPOS4 for pp collisions at $\sqrt{s} = 13$ TeV.
    }
	\label{fig:integrated_particle_ratio}
\end{figure}

The $p_{\rm T}$-integrated particle ratios, kaon- and proton-to-pion, as a function of multiplicity $\langle {\rm d} N_{\rm ch}/{\rm d} \eta \rangle$ are shown in Fig.~\ref{fig:integrated_particle_ratio} alongside published results from pp, p--Pb and Pb--Pb collisions \cite{pp_13TeV_2020, pp_7_TeV_2019, pPb_5020GeV_2014, PbPb_2760GeV_2013,  PbPb_5020GeV_2020}. The high-multiplicity data points agree well with ratios from p--Pb and Pb--Pb collisions at similar multiplicities. The measurement presents a smooth evolution with increasing  $\langle {\rm d} N_{\rm ch}/{\rm d} \eta \rangle$, indicating that the multiplicity is the driving force of the hadrochemistry, irrespective of the collision energy and colliding system. 
An increasing trend of the ${\rm K}/{\pi}$ ratio with multiplicity may indicate strangeness enhancement. In contrast, the continuous decrease of the ${\rm p}/{\pi}$ ratio with increasing multiplicity, particularly for $\langle {\rm d}N_{\rm ch}/{\rm d}\eta\rangle >\sim4$, may be consistent with the hypothesis of baryon--antibaryon annihilation in the hadronic phase, which is expected to be less significant in low-multiplicity or peripheral collisions \cite{Steinheimer-PRL_2013}. Furthermore, the $p_{\rm T}$-integrated particle ratios are compared with the corresponding predictions from the PYTHIA 8 and EPOS4 models.
In general, PYTHIA 8 CLR-BLC Mode 3 tune systematically underestimates the measured integrated ${\rm K}/{\pi}$ ratio. The other PYTHIA 8 tunes, Monash, Shoving, Ropes explain the ratio quite well, although none of the PYTHIA 8 tunes could explain the $p_{\rm T}$-differential ratio. Both PYTHIA 8 Monash and Shoving provide a qualitative
description of the ${\rm p}/{\pi}$ ratio for $\langle {\rm d} N_{\rm ch}/{\rm d} \eta \rangle>5$; however, the PYTHIA 8 Ropes and PYTHIA 8 CLR-BLC 3 could not explain the data. EPOS4 model follows the data trend of both ${\rm K}/{\pi}$ and ${\rm p}/{\pi}$ ratios and qualitatively agrees with the data results.  

\begin{figure}[h]
	\centering
	\vspace{-3 mm}
	\includegraphics[width=1\textwidth]{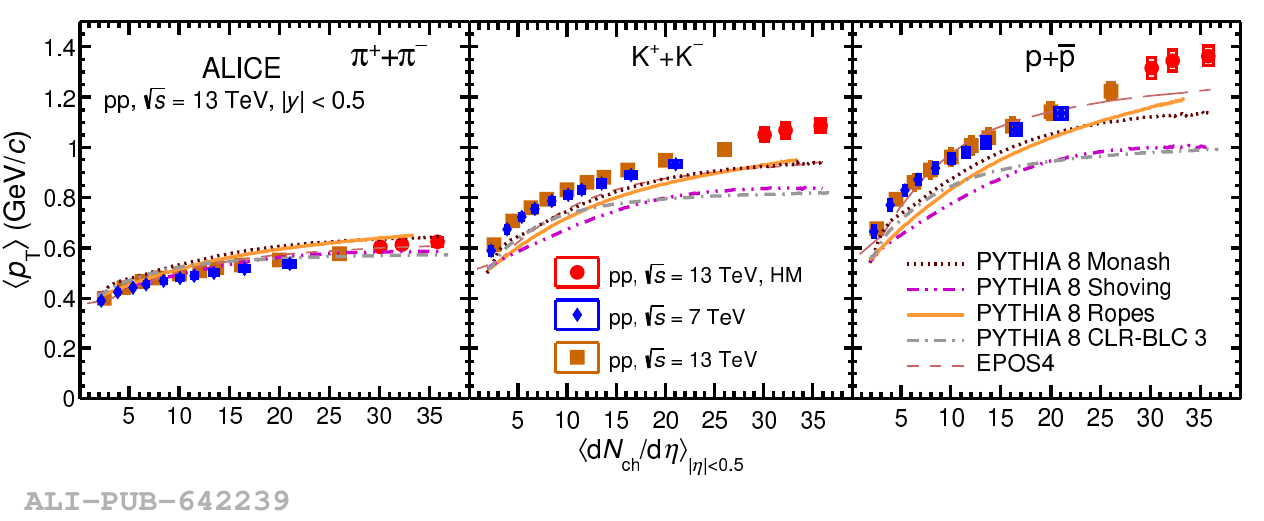}
	
	\caption{Mean transverse momentum $\langle p_{\rm T} \rangle$ of $\pi$, K, and p as a function of charged-particle multiplicity density ($\langle {\rm d} N_{\rm ch}/{\rm d} \eta \rangle$) in high-multiplicity pp collisions at $\sqrt{s}$ = 13 TeV. The results are compared with published measurements in pp collisions at $\sqrt{s}$ = 7 and 13 TeV \cite{pp_13TeV_2020, pp_7_TeV_2019}. Open boxes indicate systematic uncertainties, and bars indicate the statistical ones. Colored lines represent the predictions from different tunes of PYTHIA 8 and EPOS4 for pp collisions at $\sqrt{s}=$ 13 TeV.}    
	\label{fig:mean_pT}
\end{figure}

The measurement of the mean transverse momentum $\langle p_{\rm T} \rangle$ as a function of charged-particle multiplicity density, $\langle {\rm d} N_{\rm ch}/{\rm d} \eta \rangle$, for $\pi$, K and p is shown in Fig.~\ref{fig:mean_pT}, together with the published results of pp collisions at $\sqrt{s}$ = 7 and 13~TeV \cite{pp_13TeV_2020, pp_7_TeV_2019}. A continuous increase in $\langle p_{\rm T} \rangle$ with increasing $\langle {\rm d} N_{\rm ch}/{\rm d} \eta \rangle$ is observed, with a steeper rise for hadrons with larger masses. The present measurement confirms this trend and extends the explored $\langle {\rm d} N_{\rm ch}/{\rm d} \eta \rangle$ to higher multiplicities. Similar observations have been reported in pp \cite{CMS_pp_2012} and p--Pb \cite{pPb_5020GeV_2014} collisions at lower collision energies. The observed increase in $\langle p_{\rm T}\rangle$ with multiplicity may be indicative of stronger collective effects, such as radial flow, at higher multiplicities. A similar mass-dependent increase in $\langle p_{\rm T}\rangle$ is observed in heavy-ion collisions \cite{PbPb_5020GeV_2020}, although the magnitude of $\langle p_{\rm T}\rangle$ differs from that observed in pp collisions.

A comparison of the measured $\langle p_{\rm T} \rangle$ with the corresponding predictions from PYTHIA 8 and EPOS4 shows that all models reproduce the increasing trend of $\langle p_{\rm T} \rangle$ with multiplicity. Among the PYTHIA 8 tunes, the Shoving model provides a good description of the data across the full multiplicity range. However, all PYTHIA 8 tunes systematically underestimate the $\langle p_{\rm T} \rangle$ of kaons and protons. The $\langle p_{\rm T} \rangle$ of pions predicted by EPOS4 is consistent with the data over the full multiplicity range. For protons, EPOS4 also provides a reasonable description of the data, except for the high-multiplicity (HM) pp classes, where it underestimates the measured $\langle p_{\rm T} \rangle$. EPOS4 also underestimates the $\langle p_{\rm T} \rangle$ of kaons over the multiplicity range considered.

\section{Conclusions}
This analysis presents measurements of $\pi$, K, and p production in pp collisions at $\sqrt{s}=$13 TeV, focusing on high-multiplicity events. These measurements extend previous studies of pp collisions by increasing the average charged-particle multiplicity densities, $\langle {\rm d} N_{\rm ch}/{\rm d} \eta \rangle$, from 26.0 to 35.8, which extends the multiplicity range into the region covered by peripheral Pb–Pb collisions.
The transverse momentum spectra of the $\pi$, K and p, the ${\rm K}/{\pi}$ and ${\rm p}/{\pi}$ ratios of the $p_{\rm T}$-spectra, $p_{\rm T}$-integrated particle yield and mean transverse momentum $\langle p_{\rm T}\rangle$ are presented. The results are compared with the published data from different collision systems and energies, as well as with predictions from PYTHIA 8 (Monash, Shoving, Ropes and CLR-BLC mode 3 tunes) and EPOS4 models.

A hardening of the $p_{\rm T}$ spectra with increasing multiplicity is observed, with a more pronounced effect for particles with larger masses. The $p_{\rm T}$-differential $\rm{p}/\pi$ ratio shows an enhancement at intermediate $p_{\rm T}$. The $p_{\rm T}$-integrated yield ratios, ${\rm K}/{\pi}$ and ${\rm p}/{\pi}$, exhibit a smooth transition across different collision systems and collision energies. These observations suggest that the hadrochemistry exhibits a strong dependence on multiplicity, regardless of collision systems and collision energies. 
Furthermore, mean transverse momentum $\langle p_{\rm T}\rangle$ of the identified particles increases with charged-particle multiplicity density, $\langle {\rm d} N_{\rm ch}/{\rm d} \eta \rangle$, with a more pronounced increase for heavier particles. These features, namely the hardening of $p_{\rm T}$ spectra with multiplicity and particle masses.  In comparisons with model predictions, EPOS4 generally provides a more consistent description of the qualitative trends of the $p_{\rm T}$-dependent ${\rm K}/{\pi}$ and ${\rm p}/{\pi}$ ratios than the considered PYTHIA 8 tunes. The PYTHIA 8 Shoving tune and EPOS4 provide a good description of the ${\rm K}/{\pi}$ ratio for pions, while EPOS4 describes the measured ${\rm p}/{\pi}$ ratio reasonably well, except for the high-multiplicity (HM) classes.

\bibliography{QCD_Work-2026_bibliography}
%

\end{document}